\documentclass[epjCONF]{svjour}
\usepackage{graphics}
\usepackage[varg]{txfonts} % Times fonts
\usepackage[latin1]{inputenc}
\usepackage{natbib}

\def\apj{ApJ}\def\apjs{ApJS}\def\apjl{ApJLetters}\def\nat{Nature}\def\aap{A\&A}\def\mnras{MNRAS}
\newcommand{\ind}[1]{_{\mathrm{#1}}}

\newcommand{\refeq}[1]{(\ref{#1})}

\def\Kepler{\emph{Kepler}}

\def\numax{\nu\ind{max}}
\def\nmax{n\ind{max}}

\def\ng{n\ind{g}}

\newcommand\Tg{\Delta\Pi_1}\newcommand\tgobs{\Delta P\ind{obs}}
\def\Dnu{\Delta\nu}
\def\d01{d_{01}}

\def\Teff{T\ind{eff}}

\def\visib{V^2}
\def\visiun{\visib_1}

\def\numax{\nu\ind{max}}
\def\nmax{n\ind{max}}
\def\Dnu{\Delta\nu}
\def\Dnuobs{\Delta\nu\ind{obs}}
\def\epsobs{\varepsilon\ind{obs}}
\def\Dnuas{\Delta\nu\ind{as}}

\def\Teff{T\ind{eff}}

\newcommand\nuref{\nu\ind{ref}}
\newcommand\dnuref{\Dnu\ind{ref}}
\newcommand\Ts{T_\odot}
\newcommand\Rs{R_\odot}
\newcommand\Ms{M_\odot}
\newcommand\Rsis{R\ind{seis}}
\newcommand\Msis{M\ind{seis}}

\session-title{LIAC40}
\begin{document}
\title{Red giant seismology: observations}
\author{Beno{\^\i}t Mosser\inst{1}\fnmsep\thanks{\email{benoit.mosser@obspm.fr}}
}
\institute{LESIA, CNRS, Universit\'e Pierre et Marie Curie,
Universit\'e Denis Diderot, Observatoire de Paris, 92195 Meudon
cedex, France}
\abstract{The CoRoT and Kepler missions provide us with thousands
of red-giant light curves that allow a very precise asteroseismic
study of these objects. Before CoRoT and Kepler, the red-giant
oscillation patterns remained obscure. Now, these spectra are much
more clear and unveil many crucial interior structure properties.
For thousands of red giants, we can derive from the seismic data
precise estimates of the stellar mass and radius, the evolutionary
status of the giants (with a clear difference between clump and
RGB stars), the internal differential rotation, the mass loss, the
distance of the stars... Analysing this mass of information is
made easy by the identification of the largely homologous
red-giant oscillation patterns. For the first time, both pressure
and mixed mode oscillation patterns can be precisely depicted. The
mixed-mode analysis allows us, for instance, to probe directly the stellar core.
Fine details completing the red-giant oscillation pattern then
provide further information for a more detailed view on the
interior structure, including differential rotation. }
\maketitle
%_____________________________________________________________________________________
\section{Introduction \label{intro}}

The CNES CoRoT mission \citep{2008Sci...322..558M} and the NASA
\Kepler\ mission \citep{2010Sci...327..977B} have opened a new era
in red giant asteroseismology \citep{2009Natur.459..398D}, with
thousands of high-precision photometric light curves. This amount
of data has motivated collaborative working in dedicated working
groups, in an organisation very profitable for promoting
efficient work and impressive results
\citep[e.g.][]{2009A&A...506..465H,2010ApJ...713L.176B,2010ApJ...723.1607H,2010A&A...517A..22M}.
This paper has benefitted from all this work.

Before space-borne observation, ground-based observations have
revealed that red giants, with an outer convective envelope, show
solar-like oscillations \citep[e.g.][]{2002A&A...394L...5F}. Owing
to their low gravity, oscillations in red giants are excited at
low frequency.  Owing to their low mean density, their oscillation
pattern show frequency differences at low frequency, with the
so-called large separation of the order of a few microhertz.
Limitations due to both too short observing runs (even if the
longest lasted about two months) and a poor duty cycle have
hampered a rich output of these ground-based observations but
raised crucial questions concerning the degree of the observed
modes and the mode lifetimes. \cite{2002A&A...394L...5F}
explicitly state that ``a most important and exciting result of
[their] study is the confirmation of the possibility, suggested by
the results reported on $\alpha$ UMa and Arcturus, to observe
solar-like oscillations in stars on the red giant branch''. These
questions were not answered by observations with the
microsatellite MOST \citep[e.g.][]{2007A&A...468.1033B}, with time
series limited to one month. However, the pioneering role of these
observations was highly valuable, so that red giants were
considered as valuable asteroseimic targets. Without them, both
CoRoT and \Kepler\ should have missed an impressive harvest.

In Section \ref{scalingrel}, we first present results obtained
when considering global seismic parameters only. Such parameters allow us to perform ensemble
asteroseismology. The tools for identifying the individual
frequencies are then presented in Section \ref{pattern}. The
identification of the dipole mixed-mode pattern, not as easily
identifiable as the radial pressure mode pattern, is developed in Section
\ref{mixtes}. These mixed modes unveil unique properties of the
core. Open questions and upcoming work are presented in Section
\ref{conclusion}.

An introduction to red giant seismology, by
\cite{2011arXiv1106.5946C} and \cite{2012rgps.book...23M} for
theoretical aspects or by \cite{2011arXiv1107.1723B} in an
observational perspective, can be useful for setting the scene.

\begin{figure}
\resizebox{1.0\columnwidth}{!}{\includegraphics{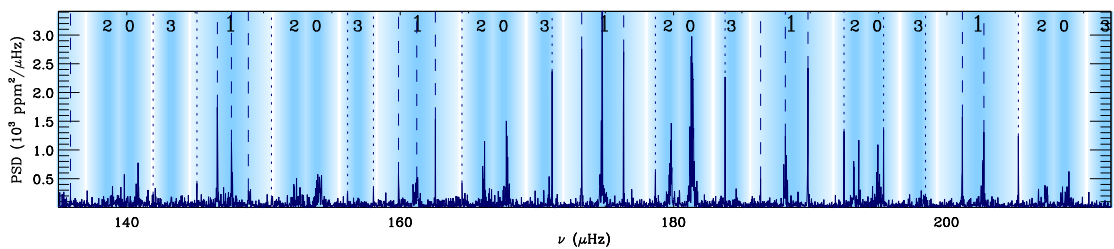}}
\caption{Power density spectrum of the star KIC 9882316, with
superimposed mode identification provided by the red giant
oscillation universal pattern. Dashed and dotted lines indicate
the position of the peaks identified as dipole mixed modes.
Pressure dominated dipole modes are located close to the positions
marked by 1. \label{exemple}}
\end{figure}

\section{Scaling relations\label{scalingrel}}

A large amount of scaling relations have been recently derived in
asteroseismology \citep[e.g.][]{2009A&A...506..465H,2010A&A...522A...1K,2011ApJ...743..143H}.
Such relations  are based on global seismic parameters used to sum up the
mean properties of a solar-like oscillation spectrum.
They allow us to perform ensemble asteroseismology, since they monitor the
evolution of various parameters for a large population of stars.

\subsection{Global seismic parameters}

Most scaling relations involve the large separation $\Dnu$ and/or the
frequency $\numax$ corresponding to the maximum oscillation
signal. The determination of these global seismic parameters can
be done diversely \citep[e.g.][]{2011A&A...525A.131H}.

Here, we use essentially the data analysis provided by the method
of \cite{2009A&A...508..877M}, called envelope autocorrelation
function (EACF). Deriving the large separation from the autocorrelation
of the time series is physically efficient, since it corresponds
to measure the delay between any oscillation signal first seen
directly, then after propagation throughout the stellar diameter
back and forth. Achieving this autocorrelation of the time series
of the oscillation signal by computing the Fourier transform of
its Fourier transform of the oscillation signal is computationally very
efficient. Considering a windowing of the spectrum, as proposed by
\cite{2009A&A...506..435R}, allows us to select a given frequency
range, or to investigate the variation of the frequency separations
with frequency, or to study independently the frequency
separations of the even and odd ridge \citep{2010AN....331..944M}.
With a filter width corresponding to the frequency range around
$\numax$ where solar-like oscillations are excited, the method
provides the mean value of the observed large separation. Last but
not least, the methodology used by the EACF method provides a test
for determining the reliability of the detection, based on the H0
hypothesis.

Scaling relation in asteroseismology is an old story, when
\cite{1917Obs....40..290E} noted that the pulsation of cepheids
are related to their mean density. This can be expressed by the
scaling relation
\begin{equation}\label{dnu_density}
\Dnu \propto \sqrt{M \over R^3}
\end{equation}
where $\Dnu$ is the mean large separation, and $M$ and $R$ are the
stellar mass and radius. $\Dnu$ is usually defined as the mean
frequency difference between consecutive radial modes
(Fig.~\ref{exemple}). In fact, this definition is misleading: frequency differences
yield the \emph{observed value} of the large separation, which is
different from the \emph{asymptotic value} that verifies Eq.
\refeq{dnu_density}. The link between the large separation and the
mean stellar density has been addressed by
\cite{2011ApJ...742L...3W} for different stellar masses and
evolutionary stages. The relation between the observed and
asymptotic values of the large separation is established by
\cite{mesure}:
\begin{equation}\label{scadnu}
\Dnuas = \left( 1 + \zeta\right) \ \Dnuobs,
\end{equation}
with
\begin{eqnarray}
% \nonumber to remove numbering (before each equation)
  \zeta &=& \displaystyle{0.57\over\nmax} \qquad\hbox{(main-sequence regime: }  \nmax \ge 15), \label{corM_MS}\\
  \zeta &=& 0.038 \qquad \ \ \hbox{(red giant regime: } \nmax \le 15), \label{corM_RGB}
\end{eqnarray}
where $\nmax = \numax / \Dnu$ measures the frequency of maximum of oscillation signal in
a dimensionless manner. The relation between $\numax$ and the
acoustic cutoff frequency $\nu\ind{c}$ proposed by
\cite{2011A&A...530A.142B} introduces the Mach number
$\mathcal{M}$ in the uppermost convective layers so that $\numax \propto
\nu\ind{c} \mathcal{M}^3$. The variation of this number with
stellar type and evolution is limited but remains unknown.

\subsection{Seismic mass and radius}

The importance of the measurements of $\Dnu$ and $\numax$ is
emphasized by their ability to provide relevant estimates of the
stellar mass and radius
\begin{equation}
  {\Rsis \over\Rs}  = \left({\numax \over \nuref}\right) \
     \left({\Dnuas \over \dnuref}\right)^{-2}
     \left({\Teff \over \Ts}\right)^{1/2}, \label{scalingR}
\end{equation}
\begin{equation}
  {\Msis\over\Ms} = \left({\numax \over \nuref}\right)^{3}
     \left({\Dnuas \over \dnuref}\right)^{-4} \left({\Teff \over \Ts}\right)^{3/2} . \label{scalingM}
\end{equation}
The reference value $\dnuref \simeq 3106\,\mu$Hz and $\nuref\simeq
138.8\,\mu$Hz have been determined by \cite{mesure}, relying on
the exact use of the second-order asymptotic expression and on the
calibration with modeled stars. Unbiased estimated of $R$ and $M$
are provided only if the asymptotic value of the large separation
is used. The use of the observed large separation induces
significant bias, of the order of 3\,\% for the radius and 6\,\%
for the mass.

Even if the calibration effort is not complete, the scaling
relation give relevant estimates. \cite{mesure} have shown that
the correct use of the scaling relations with the asymptotic large
separation provides estimates of $R$ and $M$ with uncertainties of
about 4 and 8\,\%, respectively, for low-mass stars. Uncertainties
are twice larger when $M\ge 1.3\,M_\odot$ or for red giants.

\begin{figure}\centering
\resizebox{0.65\columnwidth}{!}{\includegraphics{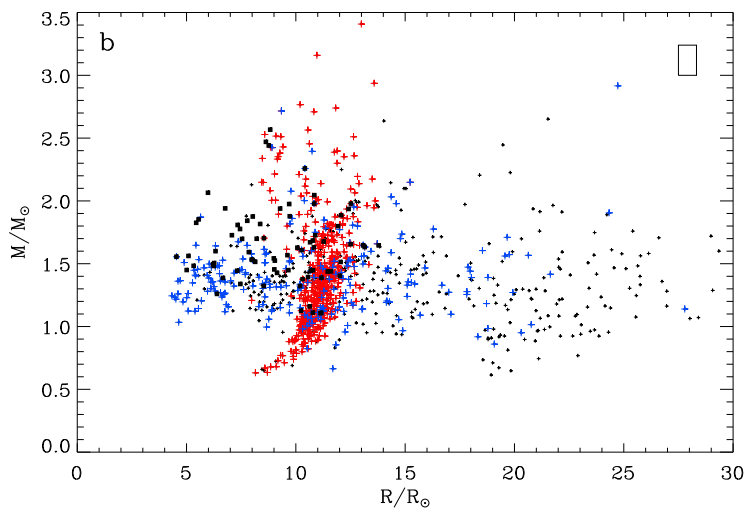}}
\caption{Asteroseismic mass as a function of the asteroseismic
radius. The color code indicates the evolutionary status; clump
stars in red, giant branch stars in blue, unknown status in dark
grey. The population of giants with low $\ell=1$ amplitude is
indicated with black squares. The rectangles in the upper right
corners indicate the mean value of the 1-$\sigma$ error bars. From
\protect\cite{2012A&A...537A..30M}.\label{correl-masse}}
\end{figure}

\subsection{Ensemble asteroseismology}

Scaling relations on global parameters allow us to perform
ensemble asteroseismology.
\begin{itemize}
    \item The radius-mass diagram puts in evidence the mass-loss occurring at
the tip of the red giant branch (RGB, Fig. \ref{correl-masse}).
The mass loss of low-mass stars is enough for reducing the mass of
their envelope to less than $0.2\,M_\odot$.
    \item \cite{2011ApJ...741..119M} have performed a comparative study of
the granulation background in giants. The parameters of the
background in the Fourier spectrum are closely related to the
parameters of the solar-like oscillation, so that a mechanism able
to partition the convective energy between oscillation and
granulation must exist. For instance \cite{2012A&A...537A..30M}
have shown that the height-to-background ratio at $\numax$ is
constant for red giants, with only a slight difference between RGB
and clump stars.
    \item Scaling relations of the oscillation amplitude were reported by different
    groups
    \citep[e.g.][for CoRoT observations]{2010A&A...517A..22M}, from
    main-sequence stars
    to red giants \citep{2011ApJ...743..143H} or using red giants in
    clusters \citep{2011ApJ...737L..10S}. Previous models have failed for reproducing the scaling
    relations. With 3D hydrodynamical models representative of the upper layers of sub- and red giant
    stars, the
acoustic mode energy supply rate computed by
\cite{2012A&A...543A.120S} shows that scaling relations of mode
amplitudes cannot be extended from main-sequence to red giants
because non-adiabatic effects for red giant stars cannot be
neglected.
\end{itemize}

\section{Frequency pattern\label{pattern}}

Any person involved in the data analysis of red giants rapidly
gets the impression that all red giant spectra are very similar.
This has to be related to the fact that red giants have
necessarily very similar interiors. Before evolving into a red
giant, the star has undergone the exhaustion of hydrogen in its
core, the contraction of its helium core, the separation of the
continuously contracting core from the continuously growing envelope,
with a thin hydrogen-burning shell at the interface, and the
growth of a large convective envelope. All these steps, mostly
governed by the properties of the hydrogen-burning shell (equation
of state, power supply rate), have erased most of the original
characteristics of the stars. After the tip of the RGB, all
low-mass red giants pass through the helium flash. As a
consequence, they gain a new opportunity to reach almost the same
interior structure, as shown by the mass-radius relation of clump
stars (Fig.~\ref{correl-masse}).

\cite{2011A&A...525L...9M} have capitalized this necessary
similarity to set up a method for measuring very precisely the
large separation and for identifying in an automated way red-giant
oscillation spectra. Assuming that these oscillations obey to a
universal pattern, they have proposed that the offset
$\varepsilon$ of the asymptotic relation
\citep{1980ApJS...43..469T} is a function of the large separation.
They have expressed the second-order term of the asymptotic
relation with a quadratic term that relates the curvature of the
ridges observed in the \'echelle diagrams:
\begin{equation}
\nu_{n,\ell} = \left[ n+{\ell\over 2}  + \varepsilon (\Dnuobs) -
d_{0\ell} (\Dnuobs) + {\alpha_\ell \over2} ( n - \nmax)^2\right]
\Dnuobs \label{tassoul_mod}
\end{equation}
We use here the subscript \emph{obs} to emphasize the difference
with the asymptotic value. The different $d_{0\ell}$ terms
indicate the small spacings of non-radial modes
\citep{2011A&A...525L...9M}. \cite{2012arXiv1209.3336M} have shown
that $\alpha_0$ is also function of the large separation. This
implies that all curvatures $\alpha_\ell$ depend on $\Dnuobs$.
This method has proven to be efficient for all red giants, with a
large separation in the range [0.4 -- 40\,$\mu$Hz], especially for oscillation spectra recorded with a low
signal-to-noise ratio.

The univocal relation between $\epsobs$ and $\Dnuobs$, updated by
\cite{2012ApJ...757..190C}, is insured if the large separation is
observed in a large frequency range. When determined in a limited
frequency range, the small difference of the offset $\varepsilon$
between RGB and clump stars allows us to determine the
evolutionary status of the giant \citep{2012A&A...541A..51K}.

\cite{mesure} have recently shown that the relation $\varepsilon (\Dnuobs )$ is an aretefact, so that the radial modes of red
giants follow the pattern:
\begin{equation}\label{redradial}
\nu_{n,0} = \left(n  + {1\over 4} + 0.037\  {\nmax^2\over n}
\right) \; \Dnuas
=
\left(n  + {1\over 4} + {18.3\over n} \left({M \over\Ms}{\Rs \over
R}{\Ts \over \Teff}\right) \right) \; \Dnuas .
\end{equation}
based on the asymptotic value $\Dnuas$ of the large separation.
This equation is fully equivalent to Eq.~\ref{tassoul_mod}, with
the relation between the asymptotic and observed values of the
large separation provided by Eq.~\ref{scadnu}.

Departures to such a regular spectrum are due to rapid structure
discontinuities. They  induce so-called glitches in the
oscillation spectrum, as due to the second ionisation of helium
\citep{2010A&A...520L...6M}. \cite{1993A&A...274..595P} have shown
that an asymptotic development can be used for addressing the
signature of such discontinuity. However, the red giant
oscillation spectrum is also much more complex, due to the
presence of other oscillation modes than pure pressure modes.

\section{Mixed modes and stellar evolution\label{mixtes}}

\subsection{Stellar evolution}

\cite{2011Sci...332..205B} have identified mixed modes in an RGB
star. Such mixed modes result from pressure waves propagating in the envelope coupled with gravity waves trapped in
the core. Due to the contraction of the inert helium core, the
Brunt-V\"ais\"al\"a frequency reaches much higher values than in
main-sequence stars, so that the coupling between the different
waves in the envelope and in the core is efficient \citep[e.g.][]{2012rgps.book...23M}. This coupling
permits the information of gravity modes to percolate to the
surface. Hence, \cite{2011Natur.471..608B} could show that the
mixed-mode frequency separation depends on the evolutionary status
of the star and allows us to distinguish helium-burning stars in
the red clump from shell hydrogen-burning stars in the RGB.
\cite{2011A&A...532A..86M} have proposed an alternative method,
based on the EACF with narrow filters centered on the dipole
modes. These first approaches only deliver the bumped period
spacing, significantly perturbed by the coupling of the pressure and gravity waves and quite
different from the period spacing $\Tg$ of gravity modes.

\subsection{Asymptotic development of the mixed mode pattern}

\begin{figure*}
\centering \resizebox{0.75\columnwidth}{!}{\includegraphics{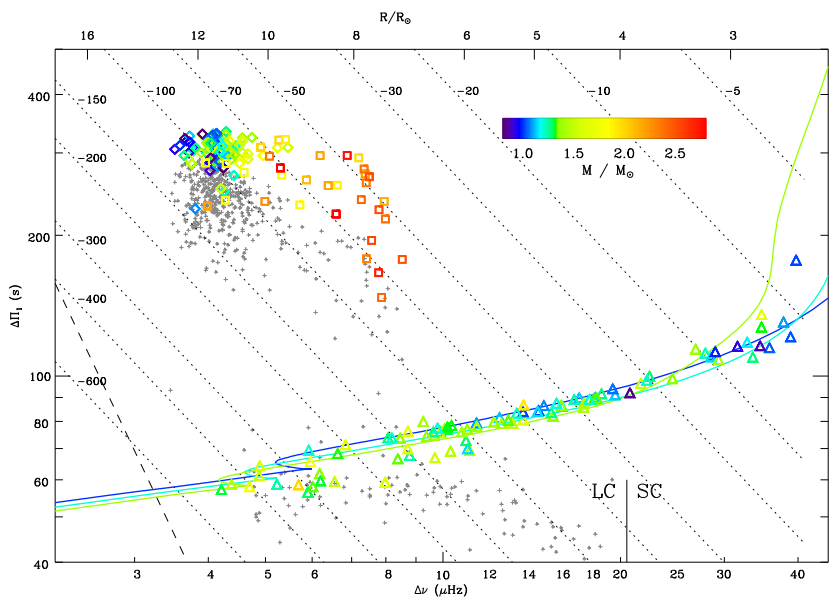}}
\caption{Gravity-mode period spacing $\Tg$ as a function of the
pressure-mode large frequency spacing $\Dnu$. Long-cadence data
(LC) have $\Dnu\le 20.4\,\mu$Hz. RGB stars are indicated by
triangles; clump stars by diamonds; secondary clump stars by
squares. Uncertainties in both parameters are smaller than the
symbol size. The seismic estimate of the mass is given by the
color code. Small gray crosses indicate the bumped periods
$\tgobs$ measured by \protect\cite{2011A&A...532A..86M}. Dotted
lines are $\ng$ isolines. The dashed line in the lower left corner
indicates the formal frequency resolution limit. The upper x-axis
gives an estimate of the stellar radius for a star whose $\numax$
is related to $\Dnu$ according to the mean scaling relation
$\numax = (\Dnu/0.28)^{1.33}$ (both frequencies in $\mu$Hz). The
solid colored lines correspond to a grid of stellar models with
masses of 1, 1.2 and $1.4\, M_\odot$, from the ZAMS to the tip of
the RGB. From \protect\cite{2012A&A...540A.143M}.
\label{spacings}}
\end{figure*}

Measuring the period spacing $\Tg$ is derived from the asymptotic
development for mixed modes exposed by \cite{2012A&A...540A.143M},
based on the method exposed by \cite{1989nos..book.....U}.
Observations of red giant with a large number of dipole mixed
modes give rise to this development. The mixed-mode frequencies
related to the pure pressure dipole mode of radial order $n$ are
solutions of the implicit equation:
\begin{equation}
\nu = \nu_{n,\ell=1} + {\Dnu \over \pi} \arctan %
\left[%
 q \tan \pi \left( {1 \over \Tg \nu} -
 \varepsilon\ind{g}
\right) \right] . \label{implicite2}
\end{equation}
where $\nu_{n,\ell=1}$ is the pure pressure mode frequency
previously determined, $q$ is a dimensionless coupling factor,
$\Tg$ is the period spacing of pure gravity modes and
$\varepsilon\ind{g}$ is a constant fixed to 0. For each pressure
radial order $n$, one obtains $\mathcal{N}+1$ solutions, with
$\mathcal{N} \simeq \Dnu\,\Tg^{-1}\numax^{-2}$. The value of $\Tg$
is derived from a least-squares fit of the observed values to the
asymptotic solution. As shown by \cite{2012A&A...540A.143M}, the
observation of high gravity mode orders insures a precise
description of the mixed-mode pattern with the asymptotic
development. As a result, the period $\Tg$ can be determined with
a high accuracy (Fig.~\ref{spacings}). This is highly valuable for
directly characterizing the stellar cores
\citep{2012rgps.book...23M}.

For low-mass stars on the RGB, the close relationship between the
large separation $\Dnu$ and the period spacing $\Tg$ emphasizes
the homology of red giants (Fig.~\ref{spacings}). This underlines the fact that the
properties of the stellar envelope are completely governed by the
properties of the helium core and its hydrogen-burning shell.

\begin{figure}
\centering
%\resizebox{0.75\columnwidth}{!}{\includegraphics{multrot009574650.ps}}
\resizebox{\columnwidth}{!}{\includegraphics{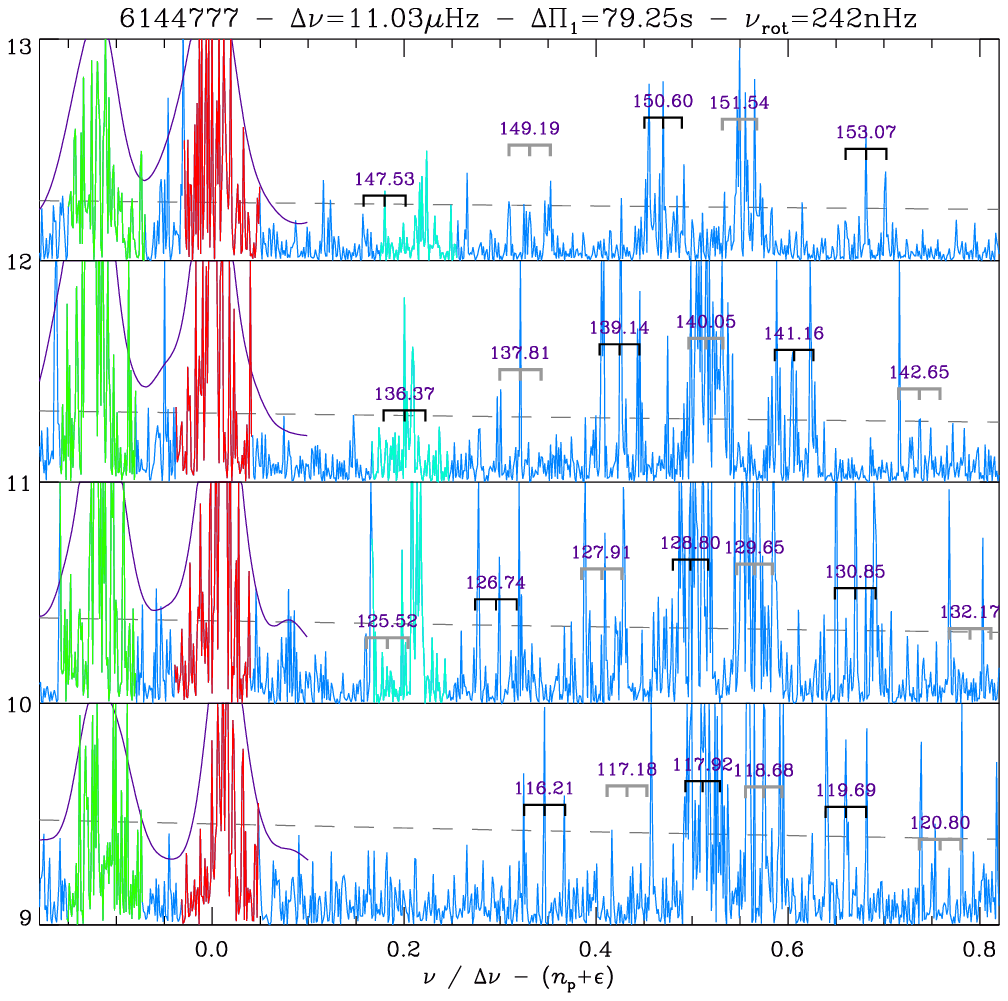}\qquad
\includegraphics{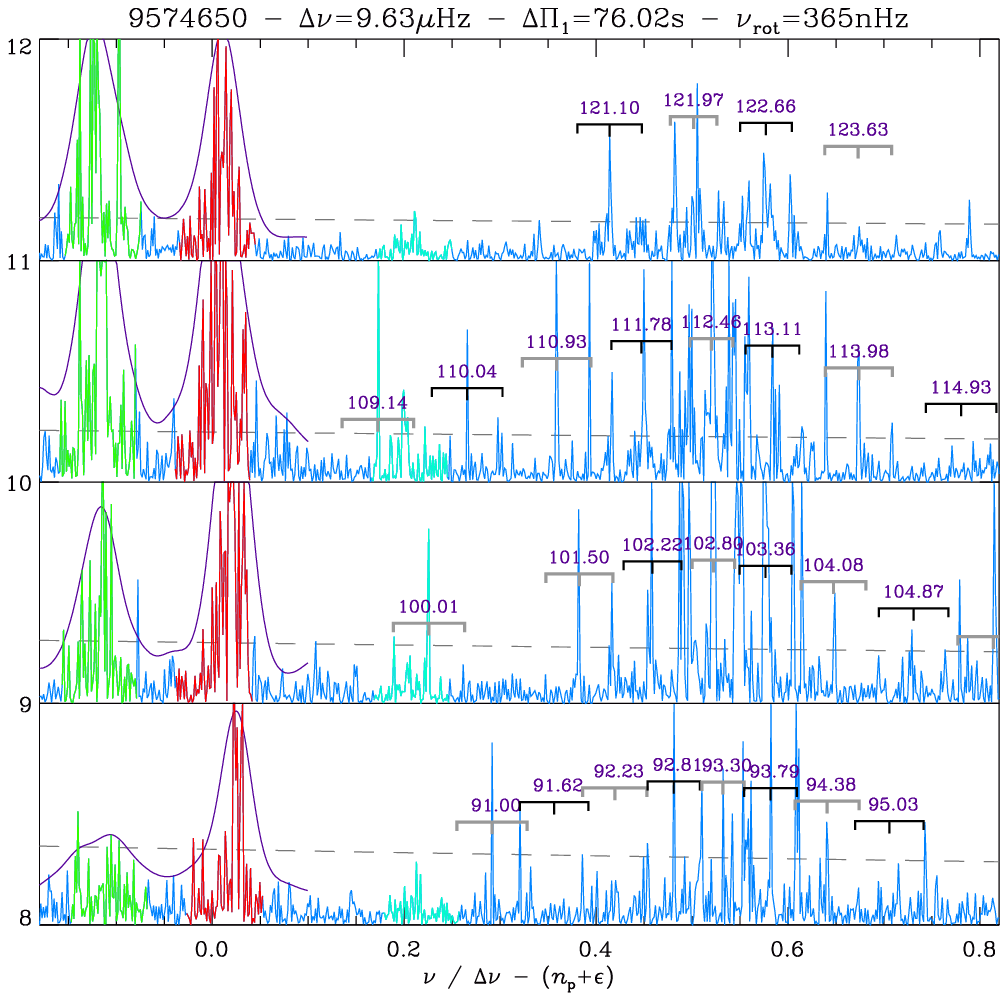}} \caption{Zoom on the
rotational splittings of the mixed modes in the giants KIC 6144777
and  9574650, in an \'echelle diagram as a function of the reduced
frequency $\nu/\Dnu - (n+\varepsilon)$. At low frequency,
multiplets are overlapping. Radial and quadrupole modes, in red
and green respectively, are located around the dimensionless
abscissae 0 and $-0.12$. The dashed lines indicates the mean value
of the background multiplied by 8. From
\protect\cite{2012arXiv1209.3336M}. \label{fig-rotation}}
\end{figure}

\subsection{Rotational splittings\label{rotational}}

\cite{2012Natur.481...55B} have shown that gravity-dominated mixed
modes revealed the core rotation in red giant. They analysed the
rotational splittings of three red giant oscillation spectra, in
the early stages of the RGB. These splittings reveal a significant
differential rotation, with a core rotating at least ten times
faster than the surface.

\begin{figure}
\centering
\resizebox{0.75\columnwidth}{!}{%
\includegraphics{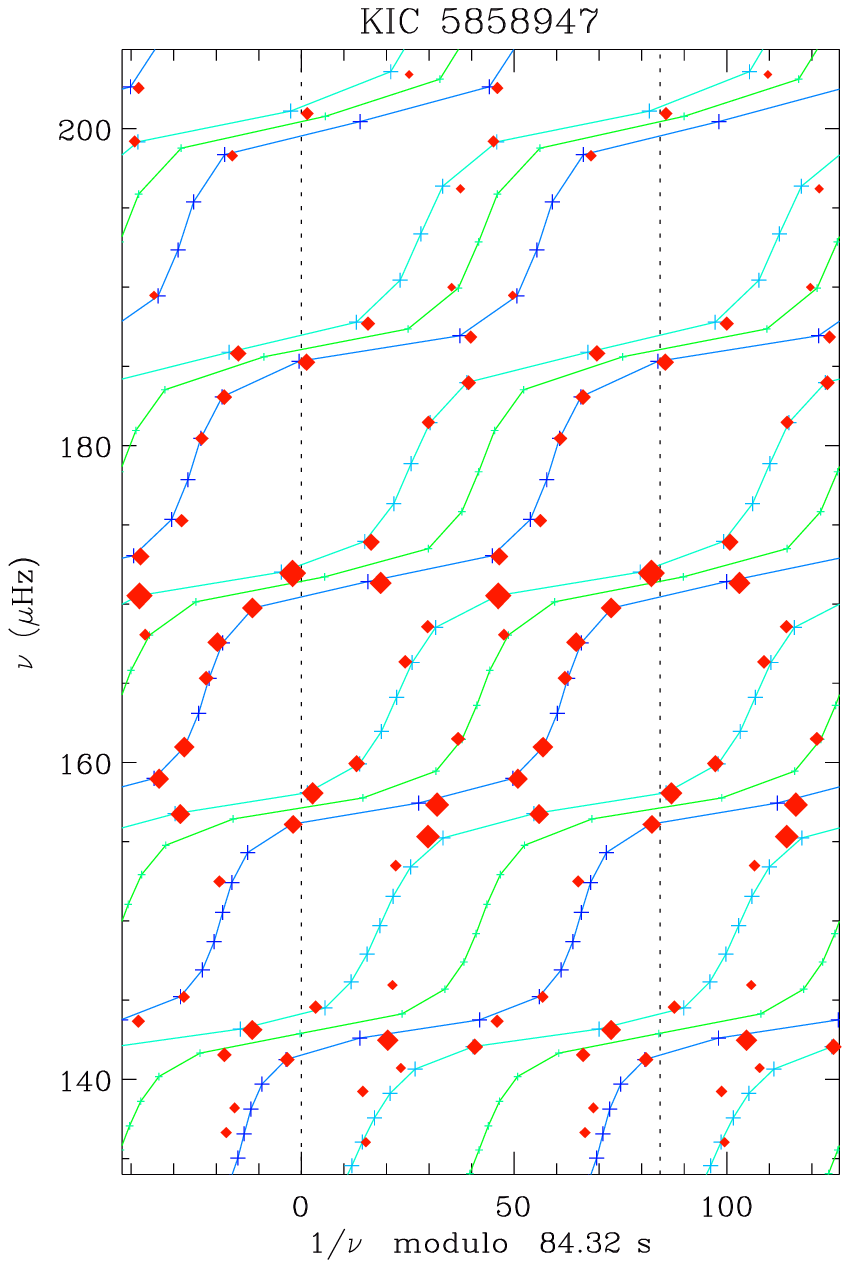}\includegraphics{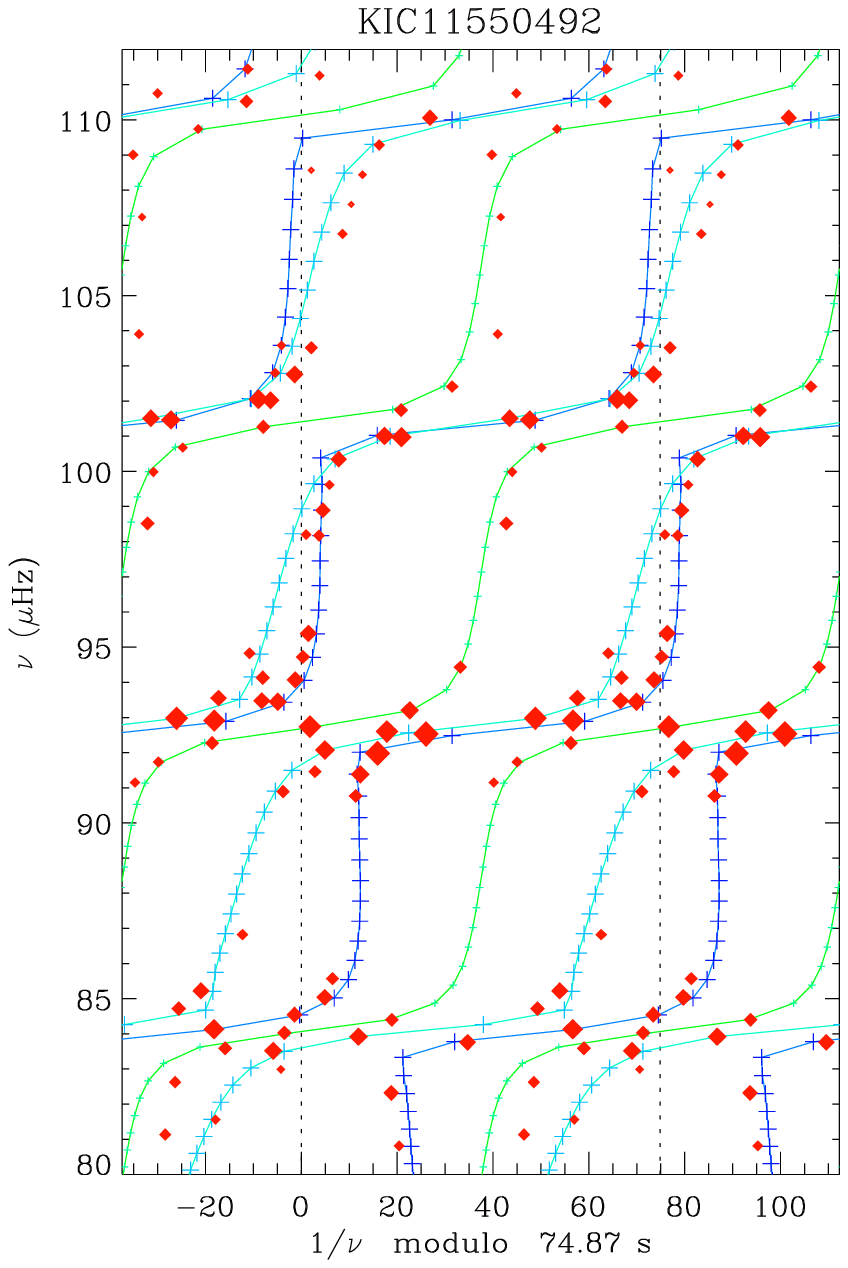} }
\caption{Gravity \'echelle diagrams of the two RGB stars KIC
5858947 and 11550492. The x-axis is the period $1/\nu$ modulo the
gravity spacing $\Tg$; for clarity, the range has been extended
from $-0.5$ to 1.5 $\Tg$. The size of the selected observed mixed
modes (red diamonds) indicates their height. Plusses give the
expected location of the mixed modes, with $m=-1$ in light blue,
$m=0$ in green and $m=+1$ in dark blue. }
\label{rotation}       % Give a unique label
\end{figure}

\cite{2012arXiv1209.3336M} have developed a method for analysing
rotation splittings in an automated way, based on the EACF
function with ultra-narrow filters. This method has provided
splittings in more than 260 red giants observed with \Kepler. A
direct identification of the rotational splittings, provided by
the method proposed by \cite{2012A&A...540A.143M}, was also used
for more than 100 red giants (Fig.~\ref{fig-rotation},
\ref{rotation}). Under the hypothesis that a linear analysis can
provide the mean core rotation from the rotational splittings of
the gravity-dominated mixed modes, the evolution of this mean core
rotation indicates a significant spin down of the core rotation
occurs in red giants. This spin down, observed on the RGB but
much more marked for clump stars, requires an significant angular momentum transport between the different regions of the star.

\begin{table}
  \centering
  \caption{Asymptotic fit}\label{fit}
  \begin{tabular}{lp{1.6cm}llp{1.9cm}}
    \hline
    Modes & Method & \multicolumn{2}{c}{Parameters} & Remark\\
          &        &   First step    & Refined step \\
    \hline
    Pressure modes& Universal  pattern & Large separation $\Dnu$&
    & {$\varepsilon = \varepsilon( \Dnu)$ \par $d_{01} = d_{01} ( \Dnu)$} \\
                 &                  & & Glitch  $\delta\varepsilon$  &  $|\delta\varepsilon|\le 0.02$ \\
                 \hline
    Mixed modes & Asymptotic & Period spacing $\Tg$ &  &$\varepsilon\ind{g}= 0$  \\
                &                       & Coupling factor $q$\\
                & & & $\varepsilon\ind{g}\ne 0$\\
                \hline
    Rotational splittings & Empirical  & Core splittings $\delta\nu\ind{rot}$ \\
                          &  & & 2 parameters\\
    \hline
  \end{tabular}
\end{table}

\section{From asteroseismic observations to stellar physics\label{conclusion}}

The analysis of the thousands of red giant oscillation spectra has
just started. The description of these spectra with the
combination of the universal red giant oscillation pattern, the
asymptotic development of mixed modes and an empirical description
of the rotational splittings has proven to be fruitful. As shown
by Table~\ref{fit}, four parameters are enough to identify all modes. Refined fits are obtained with eight free
parameters, to be compared to the number of fitted modes (in the
range 40 - 140) and to the complexity of some spectra
(Fig.~\ref{rotation}).

Undoubtedly, the high-quality asteroseismic constrains, especially
those sounding the stellar core, is promoting large progress in
stellar physics.

\subsection{Standard candles}

The precise asteroseismic constraints on red giants, and
especially the precise estimate of the radius from scaling
relations, completed with the more precise determination derived
from stellar modeling, allows us to use red giants as standard
candles \citep{2009A&A...503L..21M,2012EPJWC..1905012M}. According
to Eq.~\refeq{scalingR}, this requires the use of reliabl effective
temperatures $\Teff$, determined from photometry and colour-
$\Teff$ calibrations. Luminosities and distances are derived from
dereddened apparent 2MASS magnitudes and bolometric corrections.
Combining distances with spectroscopic constraints and
asteroseismic estimates of the mass allows a detailed
characterisation of populations of giants in different regions of
the Galaxy observed by  \Kepler\ and CoRoT at large set of
galactic latitudes and longitudes
\citep{2009A&A...503L..21M,2012EPJWC..1905012M}. This topic is
more precisely developed by \cite{miglio} in these proceedings.

\subsection{Modeling}

Modeling effort has been achieved for a limited number of red
giants with seismic constraints
\citep{2010A&A...509A..73C,2010A&A...520L...6M,2011ApJ...742..120J,2011MNRAS.415.3783D,2012A&A...538A..73B}.
If not based on grid computing, this effort is time consuming, as
it allows to address the physical input in the modeling. Then,
it makes the best of the seismic constraints. In some stars, the
lifetimes of the gravity-dominated mixed modes is so long that it
yet exceeds the total duration of the observation run (31 months
at the time this article is written), so that the accuracy of the
frequency determination is equal to the frequency resolution
($\simeq 12\,$nHz), much better than the current performance of
modeling \citep{mauro}. As a consequence, future developments are very promising.

\begin{figure}
\centering
\resizebox{0.75\columnwidth}{!}{\includegraphics{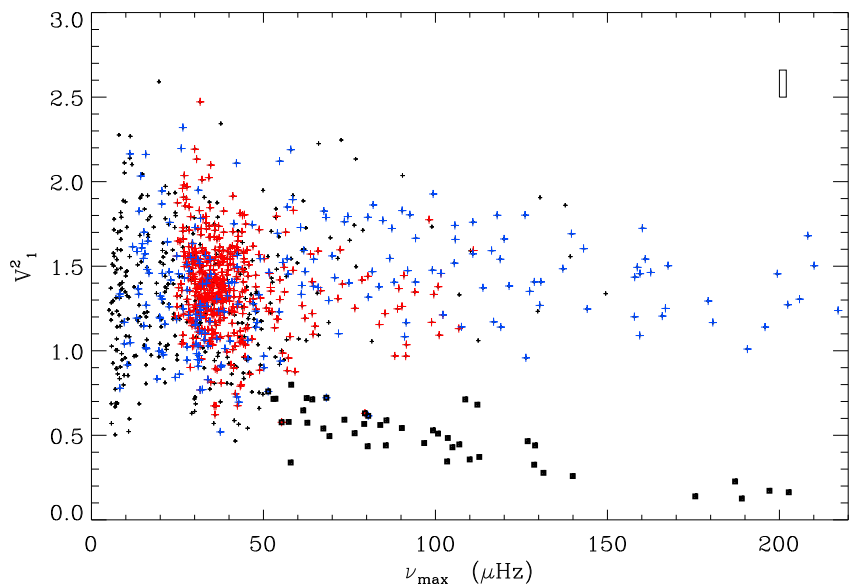}}
\caption{Visibility $\visiun$ as a function of $\numax$, with the
same color code as in Fig.~\ref{correl-masse}. Large black symbols
indicate the population of stars with very low $\visiun$ values.
\label{V1}}
\end{figure}

\subsection{Low-amplitude dipole mixed modes}

Most red giants spectra show a complex spectrum, with short-lived
pressure-dominated  and long-lived gravity-dominated mixed modes.
A family of red giants shows non-standard spectra, with depressed
dipole modes \citep{2012A&A...537A..30M}. Such red giants are
found at all evolutionary stage from the early RGB to the red
clump (Fig. \ref{V1}). The coupling between the two cavities in
the envelope and in the core certainly obeys to specific
conditions that govern such a behaviour. Clarifying the situation of these stars
will greatly help our understanding of the mixed modes in red giants.

\subsection{Upper red giant branch; asymptotic giant branch}

Red giants ascending the RGB or the AGB have such large radii that
their oscillation occur at very low frequencies, as shown by the
analysis of the upper RGB from OGLE observations
\citep{2010A&A...524A..88D}. By extrapolation of the current
results, the extension of the \Kepler\ mission can provide us with
the observation of giants with large separation as low as
0.20\,$\mu$Hz. If the scaling relations are still valid, this
corresponds to radii of about $80\,R_\odot$, maybe not enough for
investigating the tip of the RGB at all masses, but useful for
combining with OGLE results.

\subsection{Differential rotation and angular momentum transport}

The observation of the rotational splittings implies that angular
momentum is, as expected, significantly redistributed between the
different regions of the stars. A thorough analysis of this
redistribution has just started. This will take time, but we are
confident that the new constraints provided by asteroseismic
observation will be translated by theoreticians into highly
valuable information.

\end{document}